\documentclass[aps,pre,reprint,superscriptaddress,showpacs]{revtex4-1}
\usepackage{graphicx}
\usepackage{color}
\usepackage{bm}

\begin{document}

\title{Work fluctuations in a time-dependent harmonic potential: \\ rigorous results  and beyond the overdamped limit}%
\author{Chulan Kwon}%
\affiliation{Department of Physics, Myongji University, Yongin, Gyeonggi-Do 449-728, Korea}
\author{Jae Dong Noh}
\affiliation{Department of Physics, University of Seoul, Seoul 130-743,
Republic of Korea}
\affiliation{School of Physics, Korea Institute for Advanced Study, Seoul 130-722,  Korea}
\author{Hyunggyu Park}
\affiliation{School of Physics, Korea Institute for Advanced Study, Seoul 130-722,  Korea}
\date{\today}%
\begin{abstract}
We investigate the stochastic motion of a Brownian particle in the harmonic potential with a time-dependent force constant.
It may describe the motion of a colloidal particle in an optical trap where the potential well is formed by a time-dependent field.
We use the path integral formalism to solve the Langevin equation and the associated Fokker-Planck (Kramers) equation.
Rigorous relations are derived to generate the probability density function
for the time-dependent nonequilibrium work production beyond the overdamped limit. We find that the work distribution exhibits an exponential tail with a power-law prefactor, accompanied by an interesting oscillatory feature
(multiple {\em pseudo} locking-unlocking transitions) due to the inertial effect. Some exactly solvable cases are also discussed in the overdamped limit.

\end{abstract}
\maketitle

\section{Introduction}
Nonequilibrium (NEQ) fluctuation has been an important issue in the field of
statistical mechanics for the last two decades since the
first discovery of the fluctuation theorem for the entropy
production~\cite{evans,evans-searles1,gallavotti1}. Fluctuation theorems
(FT's)~\cite{gallavotti2,crooks1,kurchan1,lebowitz1,maes,sasa,seifert,
kurchan2} and the Jarzynski equality (JE)~\cite{jarzynski1,crooks2} are the
central theoretical relations governing NEQ fluctuation
phenomena, widely valid for many NEQ systems, deterministic or
stochastic, thermostatted with heat baths. It basically deals with a large
fluctuation around the average measurement in theory and experiment with considerable
contribution from rare events. Such phenomena become
dominant for a system with small degrees of freedom.
There have been extensive studies of small experimental systems such
as a microscopic bead dragging in a viscous fluid~\cite{wang}, a single
molecule of the RNA under mechanical stretch ~\cite{hummer,liphardt},
an oscillating bead under the translating center of the optical trap
~\cite{trepagnier}, the circuit of an electric dipole in electric potential
bias~\cite{garnier}, an ultra light metallic wire under torsion~\cite{douarche1}, etc.

External bias was considered as a typical underlying mechanism for NEQ systems such as the nano circuit device with potential bias~\cite{garnier,joubaud1}, the harmonic
oscillator under constant torque applied~\cite{douarche1,douarche2,joubaud2}, one dimensional lattice gas in contact at boundaries with different heat or particle baths~\cite{bodineau,derrida}.
A nonconservative force was also recognized as a source for the  entropy
production~\cite{kurchan1} such as in a nano heat engine in contact with multiple
reservoirs for a circulating current in high dimensional
systems~\cite{kwon,filliger,kwon-park-noh}.  Non-Markovian nature caused by memory effect
or colored noise is another source for NEQ~\cite{speck-seifert,puglisi1}.
In these examples, the system reaches a NEQ steady state (NESS) after a transient period,
where a persistent nonzero current, directed or circulated, generates
the incessant work production. The probability distribution function (PDF) of the work
production exhibits an exponential decay with a power-law prefactor in the rare-event
region~\cite{kwon-park-noh}, along with interesting unusual features such as
initial condition dependency of the large deviation function~\cite{farago,visco,puglisi2,jlee}
and multiple dynamic transitions in reaching the NESS~\cite{kwon-park-noh,noh2}.

On the other hand, a time-dependent perturbation on external parameters such
as electric field, magnetic field, volume, force constant, etc.~generates a
genuine time-dependent NEQ state where the system never maintains the NESS. For example, the stochastic motion of a Brownian particle was studied in the harmonic potential moving with a constant velocity ({\em sliding} parabola potential)~\cite{mazonka-jarzynski,zon,zon2,zon3,taniguchi,cohen}, and also in the harmonic potential with a time-dependent force constant ({\em breathing} parabola potential)~\cite{engel,nickelsen,speck,maass}.
In these cases, the work PDF also shows an exponential decay with a power-law prefactor in the rare-event region,
along with a time-dependent characteristic value for the work production determining the exponential decay shape.
Most of previous studies considered the overdamped
limit, partly because the experimental situation for a colloidal particle in
a harmonic trap can be well approximated in the overdamped limit and also partly
because the analytical treatment is much easier~\cite{mazonka-jarzynski,zon,zon2,zon3,taniguchi,cohen,engel,nickelsen,speck,maass}.

In this paper, we generalize these results beyond the overdamped limit ({\em underdamped} case) for a Brownian particle in a breathing parabola potential with the momentum variable kept intact. We focus on the inertial effect on the time-dependent characteristic
value for the work production. Our model may also serve as a soft-wall version of the box
expansion or compression with a single Brownian particle inside, in contact with
a thermal reservoir~\cite{lua}. The experimental setup is also feasible: In a molecular tweezer or an optical trap experiment, the potential well can be approximated by the harmonic potential. The shape of the potential well, so the force constant of the harmonic potential is set to vary with a time-dependent external field.

The stochastic motion is described by the Langevin equation and the corresponding Fokker-Planck (Kramers) equation. We use the path integral formalism to derive
rigorous relations from which the time-dependent work PDF can be easily calculated
with machine accuracy and also its cumulants at all temperatures with any time-dependent
force constant. We find an interesting oscillatory feature of the work PDF shape, solely
due to the inertial effect (absent in the overdamped limit), which resembles multiple
dynamic transitions found in the linear diffusion system~\cite{kwon-park-noh,noh2}, but
shows smooth crossovers rather than sharp transitions. Thus we call this crossover
as a {\em pseudo} locking-unlocking transition. The existence of multiple pseudo
dynamic transitions may be related to the existence of the phase-space circulating current
in the underdamped case, but full intuitive understanding calls for further investigation in future.

In section~\ref{harmonic}, we introduce the breathing harmonic potential function and discuss
the FT's. In section~\ref{sec:path-integral}, we derive the equations for
the PDF and the cumulants for the work production using the path integral
formalism. Our formalism is tested for the systems in the sudden change limit.
In section~\ref{numerical}, we present the analysis for the work
PDF and find the exponential tail with a power-law prefactor. In
section~\ref{simple}, we study the overdamped limit for exactly solvable
cases.
In section~\ref{discussion}, we discuss the main results of our work and the perspective to future works.

\section{Time-dependent Harmonic Potential}\label{harmonic}

We consider the Brownian motion of a particle in one dimension under the breathing
harmonic potential with a time-dependent force constant $k=k(t)$ and in contact with a heat bath.
The equations of motion are given by
\begin{eqnarray}
\dot{x}&=&p/m \nonumber\\
\dot{p}&=& -\gamma p/m-kx+\xi \label{eq_of_motion}
\end{eqnarray}
where $\gamma$ is a damping coefficient and $\xi$ is white noise with zero mean satisfying
$\langle \xi(t)\xi(t')\rangle=2d~\delta(t-t')$.
The diffusion coefficient $d$ is chosen to satisfy the Einstein relation,
$d=\beta^{-1}\gamma$, which guarantees the equilibrium (EQ) Boltzmann distribution
at inverse temperature $\beta$ in the steady state, if $k$ is constant in time.

The equations of motion can be rewritten as
\begin{equation}
\dot\mathbf{q}=-\mathsf{F}\cdot \mathbf{q}+\bm{\zeta}~,
\label{langevin}
\end{equation}
where $\mathbf{q}\equiv(x,p)^T$ and $\bm{\zeta}\equiv(0,\xi)^T$. Here,  the superscript $T$ denotes the transpose of a vector or a matrix.
The force matrix $\mathsf{F}$ is given by
\begin{equation}\label{Fmat}
\mathsf{F}=\left(
\begin{array}{cc}
0 & -1/m \\
k & \gamma/m
\end{array}
\right)~.
\end{equation}
The energy of a particle is given by $E(\mathbf{q};k) = \frac{p^2}{2m} +
\frac{kx^2}{2}$, which is written as $E=\frac{1}{2}\mathbf{q}^T \cdot \mathsf{H} \cdot \mathbf{q}$ with
a Hamiltonian matrix
\begin{equation}\label{Hmat}
\mathsf{H}=\left(
\begin{array}{cc}
k & 0 \\
0 & 1/m
\end{array}
\right)~.
\end{equation}

Let $P(\mathbf{q},t)$ be the probability density function for
finding a particle at state $\mathbf{q}$ and at time $t$. Then it satisfies
the Fokker-Planck equation, specially called the Kramers equation,
\begin{equation}
\frac{\partial P(\mathbf{q},t)}{\partial t}=\bm{\nabla}\cdot\left(\mathsf{F}\cdot \mathbf{q}+\mathsf{D}\cdot\bm{\nabla}\right)P(\mathbf{q},t)~,
\label{FP_eq}
\end{equation}
where $\bm{\nabla}=(\partial_x,\partial_p)^T$ and the diffusion matrix is given by
\begin{equation}
\mathsf{D}=\left(
\begin{array}{cc}
\epsilon & 0 \\
0 & d
\end{array}
\right)~,\label{singular_D}
\end{equation}
where a small positive parameter $\epsilon$ is introduced for convenience, making
possible the inversion of the diffusion matrix $\mathsf{D}$ during a formal
manipulation in the path integral formulation. In the end, we take
the $\epsilon \to 0$ limit to recover the delta function
constraint $\delta(\dot{x}-p/m)$ for position and momentum.

With $\epsilon$, position and momentum can be treated on the same footing, which gives
us the formal advantage over the usual path integral with the $\delta$-function constraint.
This approach works well. For instance, one can reproduce the expected
results for the EQ PDF when $k$ is a time-independent constant~\cite{exp1}.
In this case, the EQ Boltzmann distribution
\begin{equation}\label{B_dist}
P_{eq}(\mathbf{q};k) = \frac{1}{Z(k)} e^{-\beta E(\mathbf{q};k)}
\end{equation}
becomes the stationary solution of the Kramers equation in the limit
$\epsilon\to 0$. The partition function is given by
$Z(k) = \int d\mathbf{q}~ e^{-\beta E(\mathbf{q};k)} = (4\pi^2 m/ (\beta^2 k))^{1/2}$, so that
the free energy is given by $\mathcal{F}(k) = -\frac{1}{2\beta} \ln(4\pi^2
m/ (\beta^2 k))$.

When the force constant $k$ varies in time, the system is driven into a NEQ state.
It belongs to the Jarzynski's criterion for NEQ, where the rate of work
production is given by $\dot\mathcal{W}=\dot{k}(\partial E/\partial k)$.
Then the NEQ work $\mathcal{W}$ done on the particle
moving along a path $\mathbf{q}(\tau)$ for $0<\tau<t$ is given by
\begin{equation}
\beta{\mathcal W}[\mathbf{q}] = \beta \int_0^t d\tau \dot{k} \frac{\partial
E(\mathbf{q}(\tau);k(\tau))}{\partial k} = \frac{\beta}{2}\int_0^t d\tau \dot{k} x^2 ~.
\end{equation}
This is rewritten in a matrix form as
\begin{equation}
\beta{\mathcal W}[\mathbf{q}] = \frac{1}{2}\int_0^t d\tau \mathbf{q}^T\cdot \mathsf{\Lambda}\cdot \mathbf{q}~,
\label{work_prod}
\end{equation}
where
\begin{equation}
\mathsf{\Lambda} = \beta \dot\mathsf{H}
=\left(
\begin{array}{cc}
\beta\dot{k} & 0\\
0 & 0
\end{array}
\right)~.
\label{Lambda}
\end{equation}

The system is assumed to be initially in EQ at $\beta$ with $k(0)=k_i$ and will reach a final state with $k(t)=k_f$, which is certainly far from EQ. In this situation, the JE states
\begin{equation}
\langle e^{-\beta\mathcal{W}[\mathbf{q}]}\rangle=e^{-\beta\Delta \mathcal{F}}~,
\label{JE}
\end{equation}
where $\langle \cdots \rangle$ denotes the average over all possible paths $\mathbf{q}(\tau)$
 and $\Delta\mathcal{F}$ is the free energy difference $\mathcal{F}(k_f)-\mathcal{F}(k_i)$ at $\beta$.
The JE can be trivially derived from the Crooks relation~\cite{crooks1}
\begin{equation}
P_F(W)=e^{\beta(W-\Delta\mathcal{F})}P_R(-{W})~,
\label{crooks-FT}
\end{equation}
where $P_F({W})=\langle \delta(W-\mathcal{W}[\mathbf{q}])\rangle_F$ is the PDF for the work production $W$
  during the {\em forward} process with the change from $k_i$ to $k_f$ and vice versa $P_R(\mathcal{W})$ for the
  {\em reverse} process. The JE and the Crooks relation can be proved for a general form of energy and external perturbation for the Langevin dynamics, if the system is initially in EQ at $\beta$~\cite{kurchan2}. However, the explicit expression for $P(W)$ is not generally known. There are not many stochastic models which can be solved analytically for the PDF of fluctuating quantities. The breathing harmonic potential with a time-dependent force constant is not only analytically tractable, but can also serve an appropriate model for the potential well in an optical tweezer or trap.


\section{Path integral formalism with time-dependent force}\label{sec:path-integral}

The Fokker-Planck equation for a multivariate system with a
linear drift force, known as the high dimensional Ornstein-Uhlenbeck
process, is solvable, i.e., the time-dependent PDF $P(\mathbf{q},t)$  can be obtained
analytically~\cite{gardiner,risken}.
NEQ properties for
this process were investigated in detail from the view of the
circulating NESS current~\cite{kwon} and the violation of the
fluctuation-dissipation relation~\cite{eyink}.
Recently we have revisited this system in the light of the fluctuation
theorem~\cite{kwon-park-noh} in the case that the (non-conservative) drift
force does not vary with time.
The path integral formalism developed in that
study can be extended to the present problem with time-dependent drift
represented by the force matrix $\mathsf{F}(t)$ in
Eq.~(\ref{Fmat}) with $k=k(t)$.

To describe the NEQ fluctuations, it is convenient to introduce a path integral during time period $t$ as
\begin{eqnarray}
I(\mathbf{q}_1,\lambda; \mathbf{l}(\tau))&=&
\int d\mathbf{q}_0 P_{eq}(\mathbf{q}_0;k(0)) \int D[\mathbf{q}] \nonumber\\
&&\times e^{ -\int_0^t d\tau L(\mathbf{q},\dot\mathbf{q}) - \lambda\beta{\mathcal
W[\mathbf{q}]}+\int_0^t d\tau\mathbf{l}^T \cdot \mathbf{q} }~.
\label{path-integral}
\end{eqnarray}
The initial PDF for $\mathbf{q}_0$ is chosen to follow the EQ Boltzmann
distribution $P_{eq}(\mathbf{q}_0;k(0))$ as in Eq.~(\ref{B_dist}), and
$\int D[\mathbf{q}] (\cdots)$ denotes the integration over all possible paths
connecting $\mathbf{q}(0)=\mathbf{q}_0$ and $\mathbf{q}(t)=\mathbf{q}_1$ for $0<\tau<t$.
The Lagrangian $L$ is chosen to read
\begin{equation}\label{Lagrangian}
L(\mathbf{q},\dot\mathbf{q};\mathsf{F})=\frac{1}{4}(\dot\mathbf{q}+\mathsf{F}\cdot \mathbf{q})^T\cdot \mathsf{D}^{-1} \cdot (\dot\mathbf{q}+\mathsf{F}\cdot \mathbf{q})~.
\end{equation}
The source term $(\int d\tau~ \mathbf{l}^T \cdot \mathbf{q})$ is introduced for
a later use. Note that the exponent of the integrand is at most quadratic in $\mathbf{q}$.
Hence the path integration can be computed exactly by Gaussian integrations.

The quantity $I$ is useful in calculating physical quantities of interest.
For example, the PDF $P(\mathbf{q},t)$ is given by~\cite{onsager}
\begin{equation}\label{P_qt}
P(\mathbf{q},t) = I(\mathbf{q},\lambda=0; \mathbf{l}(\tau)=\mathbf{0})~.
\end{equation}

The PDF for the NEQ work production can be also calculated from $I$.
First, we define a dimensionless quantity for the work as
$w=\beta W$ for simplicity and introduce its generating function
\begin{equation}
\mathcal{G}(\lambda) \equiv \langle e^{-\lambda\beta \mathcal{W}}\rangle =
\int dw e^{-\lambda w} P(w)  \ ,
\label{GG}
\end{equation}
which can be obtained as
\begin{equation}\label{work_generating_lambda}
{\mathcal G}(\lambda) = \int d\mathbf{q}~ I(\mathbf{q},\lambda; \mathbf{l}(\tau)=\mathbf{0})~.
\label{gen_work}
\end{equation}
Note that the JE, $\mathcal{G}(1)=\exp [{-\beta\Delta \mathcal{F}}]$,
can be proven explicitly in this path integral formalism as well as
the generalized Crooks relation as $\mathcal{G}_F(\lambda)/\mathcal{G}_R(1-\lambda)
=\exp [{-\beta\Delta \mathcal{F}}]$ where $F$ ($R$) denotes the
forward (reverse) process.
The PDF for the dimensionless work $w$ is then obtained by the inverse
Fourier transformation as
\begin{equation}
P(w) = \int \frac{d\lambda}{2\pi} e^{i\lambda w} {\mathcal G}(i\lambda)~.
\label{IFT}
\end{equation}

For an arbitrary functional ${\mathcal A}[\mathbf{q}(\tau)]$, one can
also calculate its ensemble-averaged value from $I$.
Defining the cumulant generating function as
\begin{equation}\label{Z_mom}
{\mathcal Z}[\mathbf{l}(\tau)]= \int d\mathbf{q}~ I(\mathbf{q},\lambda=0;
\mathbf{l}(\tau)) \ ,
\end{equation}
one finds that
\begin{equation}
\langle{\mathcal A}[\mathbf{q}]\rangle =\left.
{\mathcal A}\left[\frac{\delta}{\delta \mathbf{l}(\tau)}\right]
\mathcal{Z}[\mathbf{l}(\tau)]\right|_{\mathbf{l}\to \mathbf{0}}~.
\label{expectation-value}
\end{equation}
We will use this relation to calculate the cumulants of the work.

The path integral, Eq.~(\ref{path-integral}), can be evaluated by using
the methods developed in our recent
study~\cite{kwon-park-noh}. Here, we will present the results without showing
detailed calculation steps.

\subsection{Probability distribution function}
The PDF $P(\mathbf{q},t)$ is given by
\begin{equation}
P(\mathbf{q},t)=\left|\det [2\pi \mathsf{A}^{-1}(t)]\right|^{-1/2} e^{-\frac{1}{2}\mathbf{q}^T\cdot \mathsf{A}(t)\cdot \mathbf{q}}~,
\label{time-dep-prob}
\end{equation}
where the kernel  $\mathsf{A}(t)$ is a symmetric matrix, satisfying the differential equation as
\begin{equation}
\frac{d\mathsf{A}^{-1}}{dt} = 2\mathsf{D} - \left[\mathsf{F}(t)\mathsf{A}^{-1} + \mathsf{A}^{-1} \mathsf{F}^T(t)\right]~.
\end{equation}
The formal solution is given by
\begin{eqnarray}
\mathsf{A}^{-1}(t)&=&2\int_0^{t} d\tau~ \mathsf{U}(t;t-\tau)\mathsf{DU}^T(t;t-\tau)\nonumber\\
&&+ \mathsf{U}(t;0)\mathsf{A}^{-1}(0)\mathsf{U}^T(t;0)
\label{inverse-A-zero-lambda}
\end{eqnarray}
with the initial condition $\mathsf{A}(0)=\beta \mathsf{H}(0)$.
Here the evolution operator $\mathsf{U}$ is given by
\begin{equation}
\mathsf{U}(t;t')=\left[e^{-\int_{t'}^t d\tau \mathsf{F}(\tau)}\right]_{TO}~,
\label{TO}
\end{equation}
where the subscript denotes the time-ordered product, and satisfies the
differential equation
\begin{equation}\label{diff_U}
\frac{\partial}{\partial t} \mathsf{U}(t;t') = - \mathsf{F}(t) \mathsf{U}(t;t')
\end{equation}
with $\mathsf{U}(t';t')=\mathsf{I}$ (the identity matrix).

In the absence of noises ($\mathsf{D}=\mathsf{0}$), $\mathsf{U}(t;t')$ describes the deterministic evolution
by $\mathbf{q}(t)=\mathsf{U}(t;t')~\mathbf{q}(t')$.
When the force matrix is constant in time so that $\mathsf{U}(t,t') = e^{-(t-t')\mathsf{F}}$,
one can do the integral in Eq.~(\ref{inverse-A-zero-lambda}) and find the explicit solution for $\mathsf{A}(t)$ (see Eq.~(21) in~\cite{kwon-park-noh}). For a general time-dependent $\mathsf{F}(t)$, it
is difficult to treat $\mathsf{U}$ analytically.
However, Eqs.~(\ref{inverse-A-zero-lambda}) and (\ref{diff_U}) can be solved
very precisely by numerical integrations.

\subsection{Work distribution function}

The generating function for the work distribution in Eq.~(\ref{gen_work})
involves the integration of the quantity $I$ with nonzero $\lambda$. The
work $\mathcal{W}[\mathbf{q}]$ coupled to $\lambda$ is also quadratic in $\mathbf{q}$
(see Eq.~(\ref{work_prod})), hence the integration can be performed in the same way
as was done for the probability distribution. After some algebra, one can derive
\begin{equation}
\ln \mathcal{G}(\lambda)=
-\frac{\lambda}{2}\int_0^t d\tau~ \mathrm{Tr}\left[\tilde\mathsf{A}^{-1}(\tau;\lambda)\mathsf{\Lambda}(\tau)\right]~,
\label{trace_reduction}
\end{equation}
where $\mathsf{\Lambda}=\beta \dot\mathsf{H}$ in Eq.~(\ref{Lambda}) and
$\tilde\mathsf{A}(\tau;\lambda)$ is the modified kernel due to
$-\lambda\beta \mathcal{W}$ in Eq.~(\ref{path-integral}).
It is found to satisfy the nonlinear differential equation
\begin{equation}
\frac{d\tilde\mathsf{A}^{-1}}{d\tau}=2\mathsf{D}-(\mathsf{F}\tilde\mathsf{A}^{-1}+\tilde\mathsf{A}^{-1}\mathsf{F}^T)
-\lambda\tilde\mathsf{A}^{-1}\mathsf{\Lambda} \tilde\mathsf{A}^{-1}
\label{dAIdt}
\end{equation}
where the initial condition is given by $\tilde\mathsf{A}(0;\lambda) = \beta \mathsf{H}(0)$.

This nonlinear differential equation can be solved easily for
$\lambda=0$ and $1$. The solution is $\tilde{\mathsf{A}}(\tau;0)=\mathsf{A}(\tau)$ in Eq.~(\ref{inverse-A-zero-lambda}), while $\tilde{\mathsf{A}}(\tau;1)=\beta \mathsf{H}(\tau)$.
Interestingly, $\tilde{\mathsf{A}}(\tau;1)$
corresponds to the kernel for $P(\mathbf{q},\tau)$ in the quasi-static process.
Inserting this into Eq.~(\ref{trace_reduction}), we find
\begin{equation}\label{G(1)}
\ln \mathcal{G}(1)=-\frac{1}{2}\int_0^t d\tau \left(\frac{\dot{k}}{k} \right)
=-\frac{1}{2}\ln\left[\frac{k(t)}{k(0)}\right]=-\beta\Delta \mathcal{F} ~,
\end{equation}
which verifies the JE.

It is not efficient to integrate Eq.~(\ref{dAIdt}) numerically, because
$\tilde\mathsf{A}$ sometimes becomes singular,
as $\tau$ increases in some range of $\lambda$.
Then, $\tilde\mathsf{A}^{-1}$ cannot be defined any more.
Therefore it is more convenient to
rewrite Eq.~(\ref{dAIdt}) in terms of $\tilde\mathsf{A}(\tau;\lambda)$ as
\begin{equation}\label{dAdt}
\frac{d\tilde\mathsf{A}}{d\tau} = \lambda\mathsf\Lambda +
(\tilde\mathsf{A}\mathsf{F} + \mathsf{F}^T \tilde\mathsf{A} ) - 2~\tilde\mathsf{A} \mathsf{D} \tilde\mathsf{A}~,
\end{equation}
along with an equivalent and more effective expression for the generating function
replacing Eq.~(\ref{trace_reduction}) as
\begin{equation}\label{Glambda}
\ln \mathcal{G}(\lambda) = \int_0^t d\tau~ \mathrm{Tr}
\left[\mathsf{F}(\tau)-\tilde\mathsf{A}(\tau;\lambda)\mathsf{D}\right] -
\frac{1}{2}\ln \frac{ \det \tilde\mathsf{A}(t;\lambda)}{\det \tilde\mathsf{A}(0;\lambda)}~.
\end{equation}
A similar result has been found in the time-independent
case~\cite{kwon-park-noh}. Equations (\ref{dAdt}) and (\ref{Glambda}) are ingredients for
numerical study of the work production distribution $P(w)$. Especially, the exponentially
decaying tail behavior of $P(w)$ is manifested by the divergence of $\mathcal{G}(\lambda)$,
which turned out to be fully captured by the singularity in the logarithmic boundary
term in Eq.~(\ref{Glambda}). Thus we will focus on the behavior of $\det \tilde\mathsf{A}(t;\lambda)$ in the next
section.

One can observe that  Eq.~(\ref{dAdt})
 becomes independent of $\beta$ if $\tilde\mathsf{A}$ is scaled by $\beta$.
 This proves that $\mathcal{G}(\lambda)$ as well as $P(w)$ is independent of $\beta$.
 Therefore, $P(W)$ is simply equal to $\beta P(w)$ with $w=\beta W$. In the weak noise
 (large $\beta$) limit~\cite{engel,nickelsen}, the tail behavior of $P(w)$ for large $|w|$
  determines exactly and fully the work distribution $P(W)$ except for a narrow central region, $|W|<\beta^{-1}$. We will come back to this issue later.

\subsection{Cumulants of work production}
The cumulant generating function in Eq.~(\ref{Z_mom})
is found as
 \begin{equation}\label{Z_l}
{\mathcal Z} [\mathbf{l}] =  e^{\frac{1}{2}\int d\tau\int  d\tau'
\mathbf{l}^T(\tau) \cdot \mathsf{\Gamma}(\tau,\tau')\cdot \mathbf{l}(\tau')}~.
\end{equation}
This form is expected because the Lagrangian is quadratic in $\mathbf{q}$ and the
source field $\mathbf{l}(\tau)$ is linearly coupled to $\mathbf{q}$.
The kernel $\mathsf{\Gamma}(\tau,\tau')$ is given as
\begin{equation}\label{Gamma}
\mathsf{\Gamma}(\tau,\tau')=\left\{
\begin{array}{ll}
\mathsf{U}(\tau,\tau')\mathsf{A}^{-1}(\tau') & , \hbox{$\tau\ge \tau'$;}\\ [2mm]
\mathsf{A}^{-1}(\tau)\mathsf{U}^T (\tau',\tau)& , \hbox{$\tau<\tau'$}
\end{array}\right.
\end{equation}
and $\mathsf{\Gamma}(\tau,\tau')= \mathsf{\Gamma}^T(\tau',\tau)$.

Using Eqs.~(\ref{expectation-value}) and (\ref{Z_l}), one can express any functional of path $\mathbf{q}(\tau)$ in terms of $\mathsf{\Gamma}(\tau,\tau')$.
For example, the first and the second cumulant of the work are given by
\begin{eqnarray*}
\langle{\mathcal W}\rangle &=&
\frac{1}{2}\int_0^t d\tau~ \textrm{Tr} \left[\dot\mathsf{H}(\tau) \mathsf{\Gamma}(\tau,\tau) \right] \\
\langle{\mathcal W}^2\rangle_c
&=&\frac{1}{2}\int_0^t d\tau\int_0^t d\tau'~ \textrm{Tr}\left[\mathsf{\Gamma}(\tau,\tau')
\dot\mathsf{H}(\tau')\mathsf{\Gamma}^T(\tau,\tau')\dot\mathsf{H}(\tau)\right]  ,
\end{eqnarray*}
where $\langle{\mathcal W}^2\rangle_c =\langle{\mathcal W}^2\rangle-\langle{\mathcal W}\rangle^2$.
Note that $\dot{H}_{11} = \dot{k}$ and $\dot{H}_{ab} = 0 $, otherwise. Then, the expressions become simpler:

\begin{eqnarray}
\langle{\mathcal W}\rangle &= &
\frac{1}{2}\int_0^t d\tau~\dot{k}A_{11}^{-1}(\tau) \\
\langle{\mathcal W}^2\rangle_c
&=&\frac{1}{2}\int_0^t d\tau\int_0^t d\tau'~ \dot{k}(\tau)\dot{k}(\tau')
\left(\Gamma_{11}(\tau,\tau')\right)^2.
\end{eqnarray}


One can also find higher-order cumulants in terms of $\mathsf{\Gamma}(\tau,\tau')$,
which are non-zero in all orders. This implies that the PDF
$P(w)$ should have a non-Gaussian form. The PDF shape will
be discussed further in the next session.

\subsection{Sudden change limit}\label{sudden}
A sudden change is a rare case where one can calculate the work PDF exactly,
even in the underdamped case.
Suppose that the particle is in EQ under the harmonic
potential with the force constant $k_i$, and the force constant
is changed abruptly to $k_f$ at time $t=0$~\cite{engel,nickelsen}.
If the particle is in a state
$\mathbf{q}=(x,p)^T$ just before the change ($t=0^-$),
its state still remains the same right after the change ($t=0^+$) as well as
the PDF $P(\mathbf{q},t)$. Only change occurs
in the potential energy, which results in the energy change
$\Delta E = E(\mathbf{q},k_f) - E(\mathbf{q},k_i) =
\frac{1}{2}(k_f-k_i)x^2$ for state $\mathbf{q}$. Then the work production
$\mathcal{W}(\mathbf{q})=\Delta E$.

As the initial distribution is given by the EQ Boltzmann distribution,
the PDF ${P}(x)=\int dp~ P_{eq}(\mathbf{q};k_i) = \sqrt{\beta k_i/(2\pi)} e^{-\beta k_i x^2/2}$.
Then, the PDF $P(w)$ of the dimensionless work $w=\beta \mathcal{W}$
can be easily derived, using
$P(w) dw =2 {P}(x) |dx|$, which yields that
\begin{equation}
P(w)=\left\{
\begin{array}{ll}
\theta(w)\sqrt{\frac{a}{\pi}}~w^{-1/2}e^{-aw}, &\hbox{$a>0$ }\\ [2mm]
\theta(-w)\sqrt{\frac{|a|}{\pi}}~|w|^{-1/2}e^{|a|w}, &\hbox{$a<0$ }
\end{array}\right.
\label{PDF_W}
\end{equation}
with $a=k_i/(k_f-k_i)$ and $\theta(w)$ is the Heaviside step function.

The generating function $\mathcal{G}$ can be easily calculated, using
Eq.~(\ref{GG}), as
\begin{equation}
\mathcal{G}(\lambda)=\left(\frac{\lambda k_f+(1-\lambda)k_i}{k_i}\right)^{-1/2}~,
\label{G_sudden}
\end{equation}
which diverges at $\lambda=k_i/(k_i-k_f)=-a$ as expected.
 The JE is also seen from ${\mathcal G}(1)=(k_f/k_i)^{-1/2}$.

Our analytic formalism in the previous subsections can also reproduce
 $\mathcal{G}(\lambda)$.
The sudden change in the potential function can be studied by considering
\begin{equation}
k(\tau)=k_i\theta(-\tau)+k_f\theta(\tau)~,
~~\dot{k}(\tau)=(k_f-k_i)\delta(\tau)~.
\end{equation}
%
Integrating Eq.~(\ref{dAdt}) from $\tau=0^-$ to $\tau=0^+$, one get
\begin{eqnarray}
\tilde{A}(0^+;\lambda)&=& \tilde{A}(0^-;\lambda)+\lambda\beta (k_f-k_i)
\left(\begin{array}{cc} 1&0\\0&0\end{array}\right) \nonumber\\
&=&\beta\left(
\begin{array}{cc}
(1-\lambda )k_i+\lambda k_f & 0\\
0 & 1/m
\end{array}\right)~,
\end{eqnarray}
where $\tilde{A}(0^-;\lambda)=\beta H(0^-)$ is used. Then, from Eq.~(\ref{Glambda}),
one can easily reproduce the  result in Eq.~(\ref{G_sudden}).
The cumulants of the work can be also easily calculated as
\begin{equation}
\langle w\rangle = \frac{k_f-k_i}{2k_i}  \qquad {\rm and} \qquad
\langle w^2\rangle_c = \frac{(k_f-k_i)^2}{2k_i^2}  \ .
\label{sudden_change_moment}
\end{equation}

\section{Analysis of work distribution }\label{numerical}

The analytic formalism developed in this paper is very useful to investigate the work distribution $P(w)$ numerically, in particular, its tail behavior, in contrast to direct numerical integration of the equations of motion where
we always face with a statistics problem, becoming serious in rare-event regions.
In this session, we
first present numerical  data from the latter method to check fluctuation relations and get some insights on the nature of the work production distribution. Then,
the tail behavior of $P(w)$ is carefully examined by the former method.

It is convenient to work
with dimensionless variables in numerical calculations. Rescaling $x= b_x
\tilde{x}$, $p=b_p \tilde{p}$, $t= b_t \tilde{t}$ with
$b_x=\sqrt{\frac{m}{\beta \gamma^2}}$, $b_p=\sqrt{\frac{m}{\beta}}$, $b_t = \frac{m}{\gamma}$,
one finds that the dimensionless variables $\tilde{x}$ and $\tilde{p}$
satisfy the same equations of motion as in Eq.~(\ref{eq_of_motion}) with
respect to the dimensionless time $\tilde{t}$ with dimensionless parameters
$\tilde{m}=\tilde{\gamma}=\tilde{\beta}=1$ and $\tilde{k}=
(\frac{m}{\gamma^2})k$. Hence, without loss of
generality, we will set $m=\gamma=\beta=1$. The only independent parameter is
the force constant $k$. We will consider a special case where $\dot{k}$ is
a time-independent constant, i.e., $k(t)=k_i (1+ \alpha t)$ for convenience.

First, we check the JE and the Crooks relation from direct
numerical integration of the time-discretized equations of motion.
We adopt a notation $X_n = X(t=t_n)$ for a time-dependent quantity
$X(t)$
where $t_n = n \Delta t~(n=0,1,2,\cdots)$ are discretized times in unit of
$\Delta t$.
Then, the equations of motion are solved from the difference equations
\begin{eqnarray*}
x_{n+1} &=& x_n + (\Delta t) p_n \\
p_{n+1} &=& p_n - (\Delta t) (p_n + k_n x_n ) + \sqrt{2 (\Delta t)}~ \eta_n,
\end{eqnarray*}
where $\eta_n$ are independent Gaussian-distributed random variables
with zero mean and unit variance. An initial configuration $\mathbf{q}_0=(x_0,p_0)$
is drawn from the EQ distribution of Eq.~(\ref{B_dist}). The
dimensionless NEQ work production $w_n = \beta \mathcal{W}_n$ up to time $t_n$ is evaluated
from the recursion relation
\begin{equation}
w_{n+1} = {w}_n + \frac{k_i\alpha}{4} (x_{n+1}^2+x_n^2)
\Delta t
\end{equation}
with ${w}_0=0$. Repeating the simulations $N_S$ times, one can
measure the PDF $P(w)$ and the generating function $\mathcal{G}(\lambda)$
numerically. Note that the work production $w$ is always positive for
$\alpha>0$ and negative for $\alpha<0$, independent of noise realizations.

\begin{figure}[t]
\includegraphics*[width=\columnwidth]{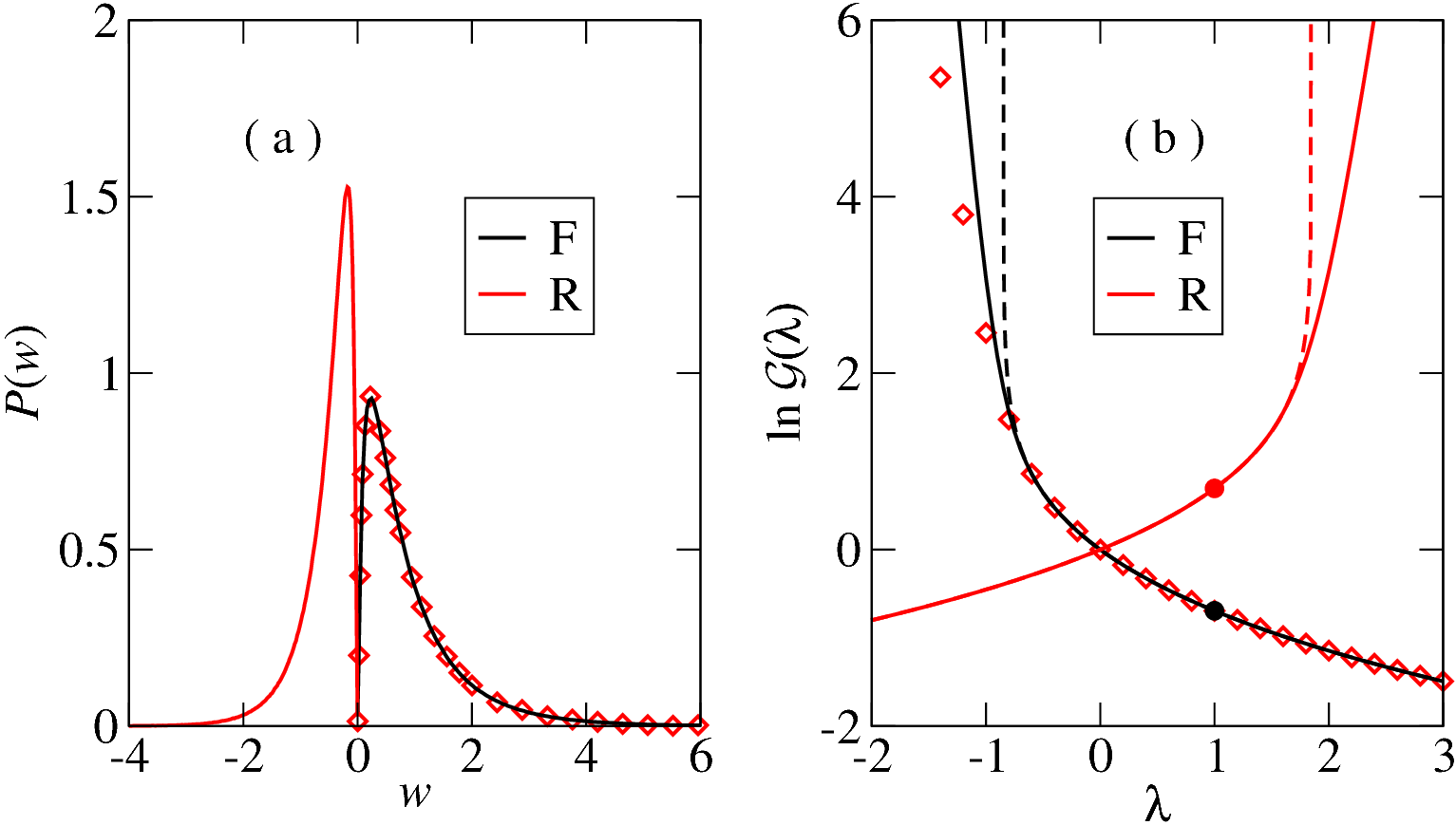}
\caption{(Color online) (a) $P_F(w)$ for the forward (F) process and $P_R(w)$ for
the reverse (R) process. Open symbols represent $e^{w-\ln 2}P_R(-w)$. (b)
$\mathcal{G}_F(\lambda)$ for the F process and $\mathcal{G}_R(\lambda)$ for
the R process. Filled symbols represent the JE points. Open symbols represent
$e^{-\ln 2}\mathcal{G}_R(1-\lambda)$. Also shown with dashed lines
are the generating functions obtained from the analytic formula in
Eq.~(\ref{Glambda}).}\label{fig1}
\end{figure}

In simulations, we take $\Delta t = 10^{-3}$ and $N_S=10^7$. The force
constant $k(t)$ is taken to vary linearly from $k_i=1$ to $k_f=4$
 and $k_i=4$ to
$k_f=1$, which will be referred to as a forward~(F) and a reverse~(R)
process, respectively. Figure~\ref{fig1}(a) shows the $P_F(w)$
for the F process till $t=3$ with $\alpha=1$ and $P_R(w)$ for the R process
with $\alpha=-1/4$.
We compare $P_F(w)$ and  $e^{w-\beta\Delta \mathcal{F}}P_R(-w)$
with $\beta\Delta\mathcal{F}=
\frac{1}{2}\ln\frac{k_f}{k_i}=\ln 2$. They seem to overlap
each other well (except for the region with very small $P(w)$), which supports the validity of the Crooks relation in Eq.~(\ref{crooks-FT}).

In order to examine the PDF in detail, we compute the generating function
$\mathcal{G}(\lambda)=\langle e^{-\lambda w}\rangle$.
These are plotted in Fig.~\ref{fig1}(b). The JE states that
$\mathcal{G}(\lambda=1) = e^{- \beta\Delta \mathcal{F}}$ where
$\beta\Delta\mathcal{F}=\ln 2$ for the F process and $-\ln 2$ for the R process.
Indeed, the numerical curves pass through the JE points. We also compare
$\mathcal{G}_F(\lambda)$ and $e^{-\beta\Delta \mathcal{F}} \mathcal{G}_R(1-\lambda)$, to check the generalized Crooks relation. For moderate values of $\lambda$, both data align along a single curve. However, there is a
slight discrepancy for large $|\lambda|$ where rare fluctuations with large
values of $|w|$ are important. This reflects a statistical uncertainty due to limited samplings.
Even with $N_S=10^7$ samples, statistics is poor for those rare
fluctuations.

Now, we utilize the analytic results in Eqs.~(\ref{dAdt}) and (\ref{Glambda})
 which are free from statistical errors, in order to determine the tail part of $P(w)$
 precisely.
In discretized times in unit of $\Delta t=10^{-3}$, the nonlinear differential
equation (\ref{dAdt}) for $\tilde\mathsf{A}(\tau;\lambda)$ is solved
with the initial condition $\tilde\mathsf{A}(0;\lambda) = \beta \mathsf{H}(0)$
and the integration in Eq.~(\ref{Glambda}) is performed numerically.
We present the numerical results for the F and R processes (dashed lines) in
Fig.~\ref{fig1}(b).
As expected, the previous simulation results deviate significantly from
our new improved numerical results in rare-event regions.
We checked that the relation $\mathcal{G}_F(\lambda) =
e^{-\beta\Delta\mathcal{F}}\mathcal{G}_R(1-\lambda)$
is satisfied perfectly well with our new numerical results at all values of $\lambda$.

\begin{figure}[t]
\includegraphics*[width=\columnwidth]{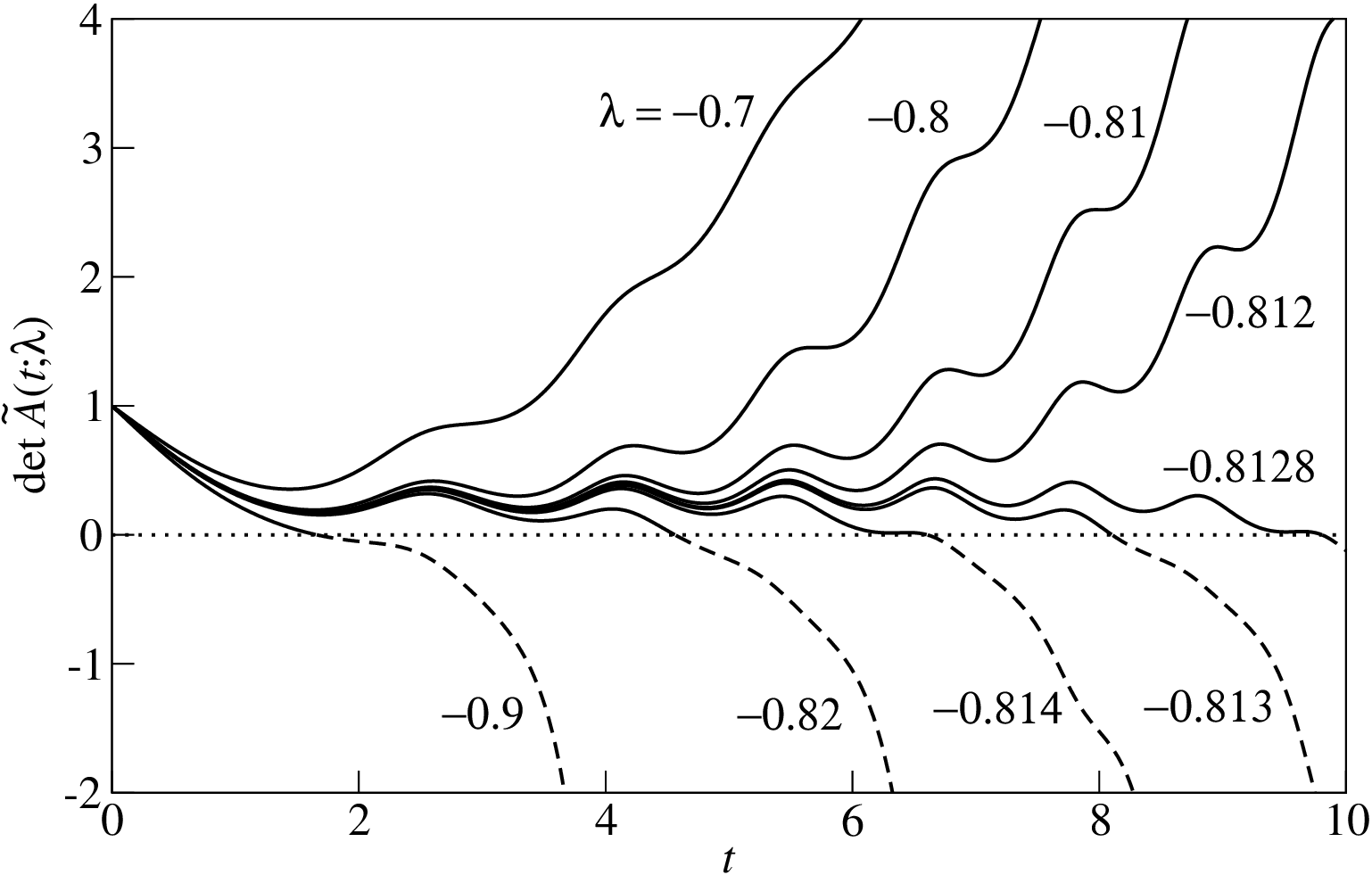}
\caption{Time evolution of $\det\tilde\mathsf{A}(t;\lambda)$ for the case with
$k(t) = k_i (1+\alpha t)$ with $k_i=1$ and $\alpha=1$ at various values of
$\lambda$.}\label{fig2}
\end{figure}

\begin{figure}[t]
\includegraphics*[width=\columnwidth]{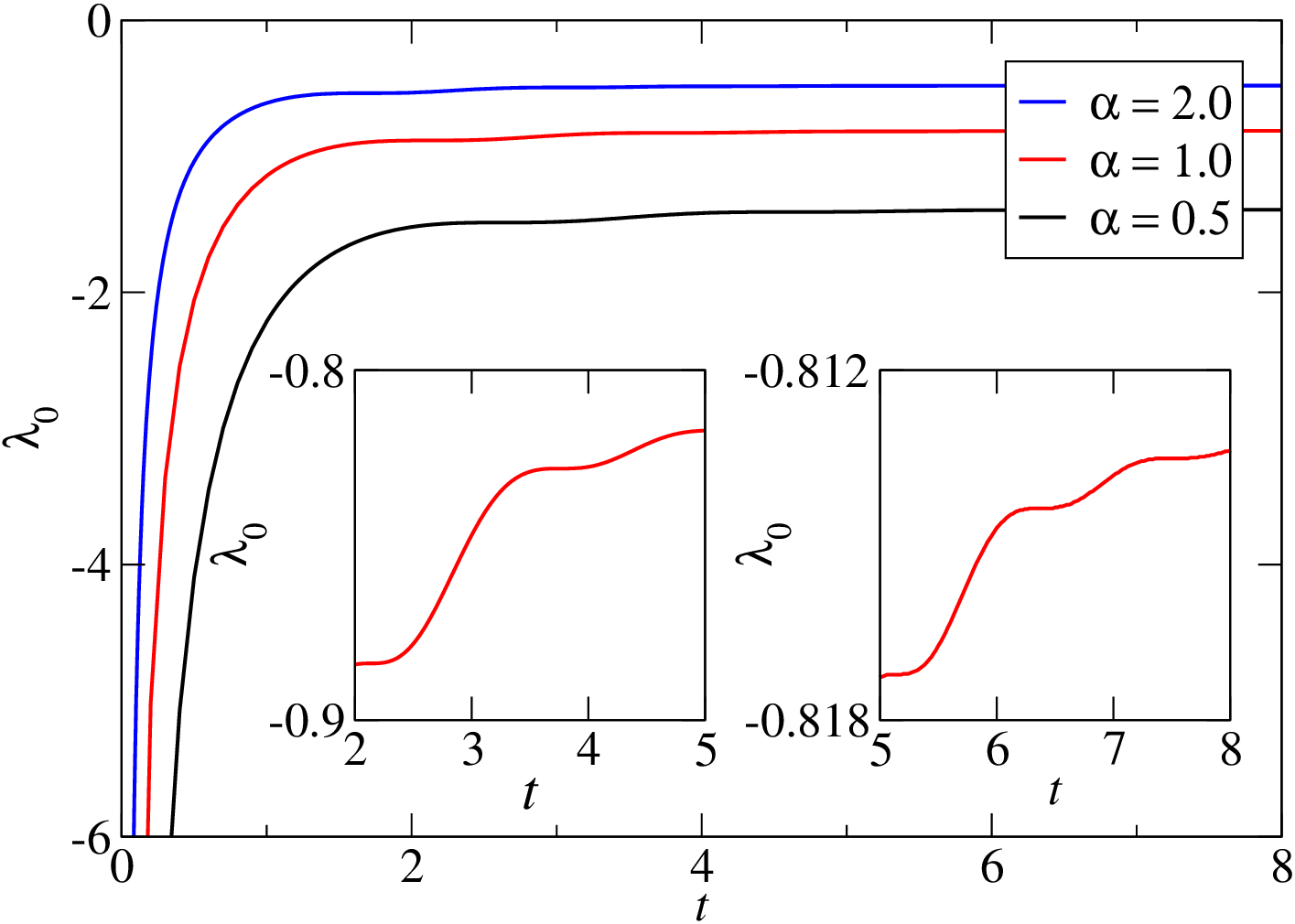}
\caption{(Color online) $t$-dependence of the threshold $\lambda_0$
for the F process with $k_i=1$ and $\alpha = 0.5, 1, 2$.
Multiple stepwise increases are observed in the insets showing the
magnification of the curve with $\alpha=1$.}
\label{fig3}
\end{figure}

In fact, our numerical data in Fig.~\ref{fig1}(b) shows that $\mathcal{G}(\lambda)$ is
divergent at a threshold $\lambda_0$ (F) and $1-\lambda_0$ (R)
with $\lambda_0 \simeq -0.84713 <0$.
The divergence occurs when
$\det{\tilde\mathsf{A}(t;\lambda)}=0$ as seen in Eq.~(\ref{Glambda}).
Figure~\ref{fig2} shows the time evolution of
$\det\tilde\mathsf{A}(t;\lambda)$ at several values of $\lambda$
in the case with $k_i=1$ and $\alpha=1$.
To a given value of $t$, $\det{\tilde\mathsf{A}}$ becomes smaller as $\lambda$ decreases
and vanishes at a threshold $\lambda_0$.
One can solve the equation $\det{\tilde{\mathsf A}(t;\lambda)}=0$ numerically
to obtain the $t$-dependent threshold $\lambda_0$.
Figure~\ref{fig3} shows the numerical results
for the system with $k_i=1$ and $\alpha=0.5, 1$, and $2$. The threshold
depends on $k_i$ and $\alpha$, and
increases monotonically and converges to a finite limiting value
$\lambda_0^\infty\simeq -1.39162, -0.81311$, and $-0.48110$ in the $t\to\infty$
limit.

The singular behavior of $\mathcal{G}(\lambda)$ reveals the asymptotic behavior of
the tail shape of $P(w)$ for large $|w|$.
Due to the generalized Crooks relation,
it suffices to consider the F  process with positive $\alpha$ (compression).
Figure~\ref{fig2} suggests that $\det\tilde{\mathsf A}(t;\lambda)$ is regular near
$\lambda=\lambda_0(t)$ so that one can write as
$\det \tilde{\mathsf A} \simeq c(\lambda-\lambda_0(t))$ with a positive constant $c$.
Then, from Eq.~(\ref{Glambda}), $\mathcal{G}(\lambda)$ diverges as
\begin{equation}\label{Gsingular}
\mathcal{G}(\lambda) \sim (\lambda-\lambda_0(t))^{-1/2} \ ,
\end{equation}
and its square-root singularity indicates that~\cite{kwon-park-noh}
\begin{equation}\label{P_tail}
P(w) \sim w^{-1/2} e^{- |\lambda_0(t)|w}
\end{equation}
for large positive $w$ with the characteristic work $w_0 \equiv 1/|\lambda_0|$.

\begin{figure}[t]
\includegraphics*[width=\columnwidth]{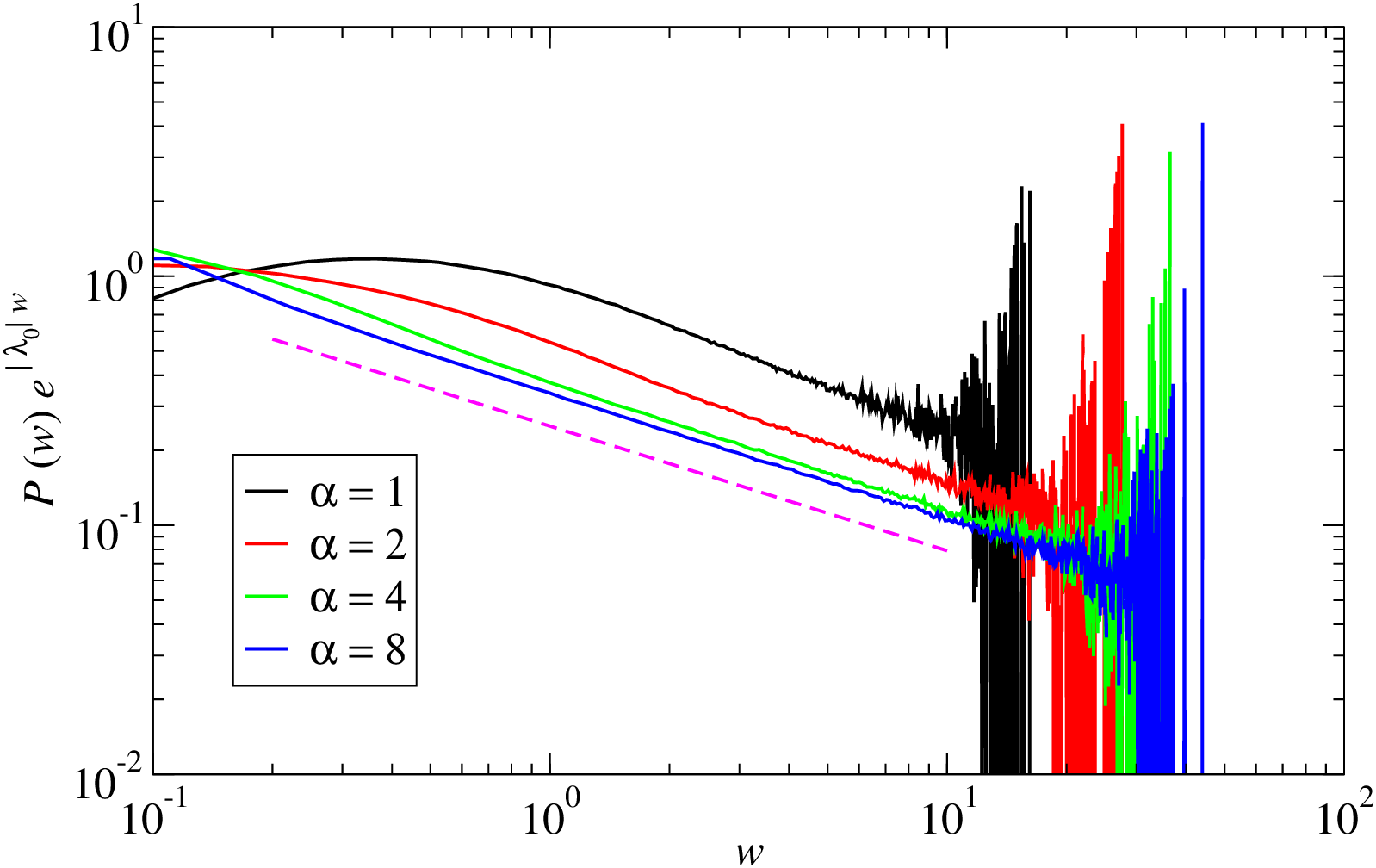}
\caption{(Color online) Rescaled PDF for the F process with
$k_i=1$, $k_f=4$, and $\alpha=1,2,4,8$.
The dashed line has a slope of $-1/2$.}\label{fig4}
\end{figure}

Our analytic formalism, combined with the numerical analysis, predicts that
there is the exponential tail in $P(w)$
with the power-law prefactor with the exponent $-1/2$. Note that
the abrupt change of $k$ (sudden change limit) also yields
the same tail, as was shown in the preceding section.
We test the tail shape by direct numerical integration of the equations of motion, using $k_i=1$,
$k_f=4$, and various $\alpha=1,2,4,8$. For each case, the threshold $\lambda_0$ is
obtained by solving the equation
$\det{\tilde{\mathsf A}}(t;\lambda_0)=0$ with fixed $t={(k_f-k_i)}/{k_i\alpha}$.
In Fig.~\ref{fig4},
the PDF $P(w)$
multiplied with $e^{|\lambda_0|w}$ follows a power-law scaling for large
$w$, which confirms the tail shape. Huge fluctuations for large $w$ are due to statistical errors
in sampling rare events.

The tail shape of $P(w)$ in Eq.~(\ref{P_tail}) is consistent with previous findings in the overdamped
limit~\cite{engel,nickelsen,speck,maass}. It is also not surprising to find that
$|\lambda_0(t)|$ decreases with $t$ because one can easily expect that
the work PDF should be more distributed (flatter) as $t$ increases.
However, the monotonic behavior of  $|\lambda_0(t)|$ is not trivially smooth, but
has an interesting repeating structure.

In Fig.~\ref{fig3}, we observe a {\em stepwise change} of $|\lambda_0(t)|$ in time,
composed of a rather fast linear change followed by quite slow plateau-type change,
which repeats itself but with decreasing size both in magnitude and time period, and
finally converges to the limiting value of $|\lambda_0^\infty|$. This implies that
the exponential tail of $P(w)$ relaxes into the limiting distribution via multiple (possibly
infinitely many) fast-slow-type relaxation dynamics. These repeated fast-slow-type
dynamics resemble multiple locking-unlocking dynamic transitions found in two-dimensional linear diffusion systems
in the overdamped limit~\cite{noh2}. However, our case shows rather smooth crossovers between
fast and slow dynamics, in contrast to sharp transitions with completely flat plateaus in
$|\lambda_0(t)|$~\cite{noh2}.
We call the stepwise changes in our case  as {\em pseudo} locking-unlocking
transitions. In the mathematical language, we can not find any $\det{\tilde\mathsf{A}}(t;\lambda)$ curve
tangential to the $t$  axis ($\det{\tilde\mathsf{A}}=0$) in Fig.~\ref{fig2}, which prohibits
a completely flat plateau, so no sharp transition is realized.

It is easy to recognize that the oscillatory feature of $\det{\tilde\mathsf{A}}$ in Fig.~\ref{fig2} evokes the stepwise change of $|\lambda_0(t)|$. First, note that
all $\det{\tilde\mathsf{A}}$ curves show oscillatory wiggles almost simultaneously
in time and the oscillation frequency grows as $t$ increases.
So we can define
a set of characteristic times $(t_1^\pm, t_2^\pm, \cdots)$ where all curves show a local
minimum ($+$) or maximum ($-$) simultaneously, at least, approximately.
The oscillatory behavior is related to the increasing frequency of the harmonic
oscillator caused by the increasing force constant $k(t)=k_i (1+\alpha t)$.

Due to this oscillatory feature of the $\det{\tilde\mathsf{A}}$ curves, one can
easily figure out that
the curves cross the $t$ axis sparsely right after $t=t_1^+$ until $t=t_1^-$,
and densely for $t_1^-<t<t_2^+$, and so on. Therefore,
$\lambda_0$ increases very fast during $0<t<t_1^+$ and very slow during
$t_1^+<t<t_1^-$, and this fast-slow relaxation dynamics repeats itself with increasing
frequency.

The underlying mechanism of these pseudo locking-unlocking transitions should
be similar to one for the sharp transitions found in the two-dimensional
linear diffusion systems~\cite{noh2}. Only differences are the nature of the rotational current, which exists here only in the phase space of $(x,p)$ and the time-dependent
external force, which acts a role of the rotational driving force as well as the
(time-dependent) anisotropic harmonic potential in the phase space.
However, we could not find a sharp dynamic transition in our model with an arbitrary choice
of parameters $(k_i, \alpha)$. Recalling what we learned in \cite{noh2}, we guess that
the anisotropy may be always small in our model, compared to the driving force magnitude,
in order to avoid a sharp transition. For full  understanding, however,
a further investigation is necessary.

For the overdamped one-dimensional case, we cannot have any rotational current, so the
oscillatory behavior is expected to be absent at all, which is confirmed rigorously
in the next session. Therefore, it can be concluded that the pseudo locking-unlocking transitions found in the underdamped case originate from the existence of the rotational
current in the phase space.

\section{Overdamped limit}\label{simple}

In the overdamped limit, the usual Fokker-Planck equation, replacing the Kramers equation, reads
\begin{equation}
\frac{\partial P(x,t)}{\partial t}=\frac{\partial}{\partial x}\left(\gamma^{-1}k(t)x+(\gamma\beta)^{-1}\frac{\partial}{\partial x}\right)P(x,t)~.
\end{equation}
Then one can use the analytic formalism developed so far for the Brownian dynamics by replacing $\mathsf{D}$ with ${(\gamma\beta)}^{-1}$, $\mathsf{F}$ with $\gamma^{-1}k$,
$\mathsf{H}$ with $k$, and $\dot\mathsf{H}$ with
$\dot{k}$. Then, the work generating function $\mathcal{G}(\lambda)$ is given by
\begin{equation}\label{lnG_overdamped}
\ln \mathcal{G}(\lambda) = \int_0^t d\tau \left(\frac{k(\tau)}{\gamma}
-\frac{\tilde{A}(\tau;\lambda)}{\gamma\beta}\right) - \frac{1}{2}\ln
\frac{\tilde{A}(t;\lambda)}{\tilde{A}(0;\lambda)} ,
\end{equation}
where the scalar quantity $\tilde{A}(t;\lambda)$ satisfies a nonlinear
differential equation
\begin{equation}\label{dAdt_overdamped}
\frac{d\tilde{A}(\tau;\lambda)}{d\tau} = (\beta\dot{k}) \lambda + 2
\frac{k}{\gamma}\tilde{A} - \frac{2}{\gamma\beta} \tilde{A}^2
\end{equation}
with the initial condition $\tilde{A}(0;\lambda) = \beta k(0) = \beta k_i$.
We will set $\gamma=\beta=1$ without loss of generality.
The overdampled limit is investigated in detail with different choices of
$k(\tau)$.

\subsection{ $k(\tau) = k_i(1 + \alpha \tau)$ }

\begin{figure}[t]
\includegraphics*[width=\columnwidth]{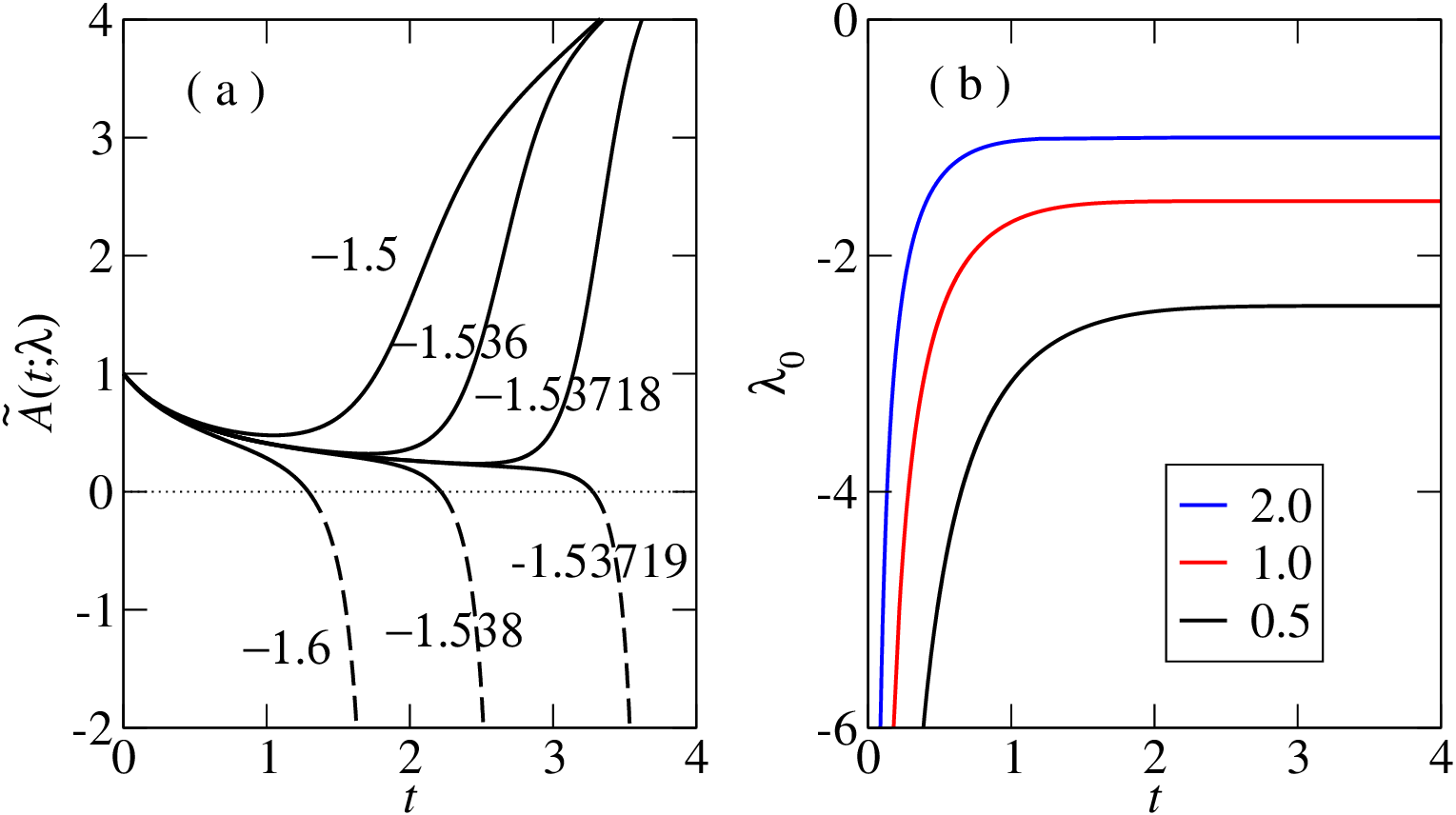}
\caption{(Color online) (a) Time evolution of $\tilde{A}(t;\lambda)$
in the overdamped case with $k_i=1$ and $\alpha=1$ at various values of $\lambda$.
(b) $t$-dependence of the threshold $\lambda_0$
for the processes with $k_i=1$ and $\alpha = 0.5, 1, 2$.}
\label{fig5}
\end{figure}

All the relevant informations are obtained from Eqs.~(\ref{lnG_overdamped})
and (\ref{dAdt_overdamped}). Unfortunately, the closed-form solution for
$\tilde{A}(t;\lambda)$ or $\mathcal{G}(\lambda)$ is not available.
However, the highly accurate numerical solution is possible, which
is shown in Fig.~\ref{fig5}(a) for $\tilde{A}(t;\lambda)$ with $k_i=1$ and $\alpha=1$. As in the
Brownian dynamics, it becomes zero at a $t$-dependent threshold
$\lambda_0$. The threshold is plotted in Fig.~\ref{fig5}(b).
>From the similar analysis for Eq.~(\ref{Gsingular}), the PDF $P(w)$ can be found to have the same tail shape as in Eq.~(\ref{P_tail}). Note that the particle dynamics does not display any oscillatory motion in the overdamped
limit, hence $\tilde{A}$ does not either as seen in
Fig~\ref{fig5}(a). Thus, the threshold $\lambda_0$ in Fig.~\ref{fig5}(b) varies in time smoothly,
showing no stepwise change at all.

Engel and Nickelsen studied the same harmonic potential problem in the overdamped limit~\cite{engel,nickelsen}. In their studies, they evaluated the path integral using the saddle point method in the low noise (large $\beta$) limit. However, as pointed out in section~\ref{sec:path-integral},
$P(w)$ is independent of $\beta$, so $P(W)$ with $w=\beta W$ can be {\em exactly} determined
from the tail behavior of $P(w)$ except for $|W|<\beta^{-1}$.

In order to calculate the cumulants of the work production, we need to evaluate the PDF kernel $A(t)$, first, using
Eqs.~(\ref{inverse-A-zero-lambda}) and (\ref{TO}). As we do not need to deal with the time ordered product,
we simply gets $U(t,t')=\exp [-\int_{t'}^{t} d\tau k(\tau)]$. Then, one can explicitly calculate the cumulants for the work production.
Using Eqs.~(\ref{inverse-A-zero-lambda}) and (\ref{Gamma}), we find
\begin{equation}
A^{-1}(t)=2\int_0^t d\tau~ e^{-2\int_{\tau}^t d\tau' k(\tau')}
+ k_i^{-1} e^{-2\int_0^t d\tau k(\tau)}~,
\label{A-inverse-overdamped}
\end{equation}
and
\begin{equation}
\Gamma(t,t')= e^{-\int_{t'}^t d\tau k(\tau)}A^{-1}(t')
\label{Gamma-overdamped}
\end{equation}
where we use $A^{-1}(0)=k_i^{-1}$.

For $k(\tau)=k_i(1+\alpha \tau)$, we find
\begin{equation}
A^{-1}(t)=k_i^{-1}e^{-2k_it-k_i\alpha t^2}g(t)~,
\end{equation}
and
\begin{equation}
\Gamma(t,t')=k_i^{-1} e^{-k_i(t+t')-\frac{1}{2}k_i\alpha(t^2+t'^2)}g(t')~,
\end{equation}
where
\begin{equation}
g(t)=1+2k_i\int_0^t d\tau~  e^{2k_i\tau+k_i\alpha\tau^2}~.
\end{equation}
The first and second cumulants are found as
\begin{eqnarray}
\langle w\rangle&=&\frac{\alpha}{2}\int_0^{t}d\tau e^{-2k_i\tau-k_i\alpha\tau^2}g(\tau)~,\\
\langle w^2\rangle_c&=&\frac{\alpha^2}{2}\int_0^td\tau\left\{\int_0^{\tau}d\tau' h(\tau,\tau')g(\tau')^2\right.\nonumber\\
&&\left.+ \int_{\tau}^{t}d\tau' h(\tau,\tau')g(\tau)^2\right\}~,
\end{eqnarray}
where
\begin{equation}
h(\tau,\tau')=e^{-2k_i(\tau+\tau')-k_i\alpha(\tau^2+\tau'^2)}~.
\end{equation}

There are two extreme cases; quasi-static and sudden processes. For the quasi-static process, one can take the limit: $\alpha\to 0$, $t\to\infty$ with
a finite value of $\alpha t=(k_f-k_i)/k_i$. Changing variable as $u=k_i\alpha\tau/(k_f-k_i)$, one can get an approximate result for the time integral.
The important ingredient for the integration is
$$
\int_a^b du~ e^{c u^2}\to \frac{1}{2c}\left(\frac{e^{cb^2}}{b}-\frac{e^{ca^2}}{a}\right)~,
$$
where $c=(k_f-k_i)^2/(k_i\alpha)\to\infty$. Then, one can get
\begin{eqnarray}
\langle w\rangle &=& \frac{1}{2}\ln\left(\frac{k_f}{k_i}\right) +
\frac{\alpha}{8}\frac{k_f^2-k_i^2}{k_i k_f^2}+\mathcal{O}(\alpha^2)~,\nonumber\\
\langle w^2 \rangle_c &=& \frac{\alpha}{4}\frac{k_f^2-k_i^2}{k_i k_f^2} +\mathcal{O}(\alpha^{2})~,
\label{quasi-static}
\end{eqnarray}
which agree with the results by Speck~\cite{speck}.
It indicates that the work distribution function is perfectly Gaussian centered around $w=\beta\Delta \mathcal{F}$
up to $\mathcal{O}(\alpha)$, and the non-Gaussianity starts to appear in $\mathcal{O}(\alpha^2)$~\cite{exp2}.
In the quasi-static limit ($\alpha\to\infty$), the work distribution function becomes a delta function,
as expected for EQ processes.

For a sudden process, one can take the opposite limit; $\alpha\to\infty$, $t\to 0$ with a finite value of $\alpha t=(k_f-k_i)/k_i$. Also using the same variable $u$, the integrand of $\int_a^b du~ e^{ cu^2}$ can be expanded
in orders of $c$ in the $c\to 0$ limit. As a result, one can get
\begin{eqnarray}
\langle w\rangle &=& \frac{k_f-k_i}{2k_i}\left(1-\frac{(k_f-k_i)^2}{3k_i\alpha}\right) +\mathcal{O}(\alpha^{-2})~,\nonumber\\
\langle w^2 \rangle_c &=& \frac{(k_f-k_i)^2}{2k_i^2}\left( 1-\frac{2(k_f-k_i)}{3\alpha}\right) +\mathcal{O}(\alpha^{-2})~.
\label{sudden_cumulant}
\end{eqnarray}
Note that $\langle w\rangle$ and $\langle w^2\rangle_c$ are finite even for an instantaneous change ($\alpha=\infty$),
which agrees with the sudden change limit for the underdamped case in Eq.~(\ref{sudden_change_moment}). It is different
from the case for the rigid wall moving with speed $v$ in the $v\to\infty$
limit, where we expect $\langle w\rangle\to 0$~\cite{lua}. The difference
is due to the distinctive situations. For the former, the collision
occurs everywhere with the harmonic potential, while for the latter the collision occurs only at the descending wall. The similarity lies in the non-trivial fluctuation around the average value.

\subsection{ $k(\tau) = k_i / ( 1 + \alpha \tau)$}

With this specific form, one can find the closed-form solution for the work
generating function $\mathcal{G}(\lambda)$. For $\alpha>0$, the harmonic potential becomes flatter
with time $\tau\ge 0$ and the work  $w$ done on the particle is always negative.
On the other hand, for $\alpha<0$, the harmonic potential becomes stiffer with time $\tau$
($0\leq \tau<1/|\alpha|$) and $w$ is always positive.

It is convenient to change variables as
\begin{eqnarray}
f_\lambda(u) &\equiv& (1+\alpha \tau) \tilde{A}(\tau;\lambda) \ , \\
u    &\equiv& \frac{1}{\alpha} \ln(1+\alpha \tau) \ .
\end{eqnarray}
The time-like variable $u$ is monotonically increasing
with $\tau$, starting from $0$ to $\infty$ for any nonzero $\alpha$.
>From Eq.~(\ref{dAdt_overdamped}), one obtains a differential equation for
$f_\lambda(u)$:
\begin{equation}
\frac{df_\lambda}{du} = -2 \left[ (f_\lambda-c)^2 + \kappa^2 \right]
\end{equation}
with $\kappa = \sqrt{\tilde{\lambda}-c^2}$,
$c = (2k_i+\alpha)/4$, and $\tilde{\lambda}= {\alpha k_i}\lambda /2$.
Note that $\kappa$ may be either positive real or pure imaginary depending on the range of
$\tilde\lambda$. In either case, the solution is given by
\begin{equation}\label{f_solution}
f_\lambda(u) = \frac{k_i \cos(2\kappa u) + ( ck_i - \tilde{\lambda})
                     \frac{\sin(2\kappa u)}{\kappa}}{
        \cos(2\kappa u) + ( k_i - c ) \frac{\sin(2\kappa u)}{\kappa}}~,
\end{equation}
with $\cos(ix) = \cosh{x}$ and $\sin(ix) = i\sinh{x}$ for any $x$.

With the solution for $f_\lambda(u)$ or equivalently for
$\tilde{A}(\tau;\lambda)$, one can obtain the work generating function
using Eq.~(\ref{lnG_overdamped}). It is useful to note that
$f_\lambda(u)=c+\frac{1}{2} \frac{d}{du} \ln [\cos(2\kappa u) + ( k_i - c ) {\sin(2\kappa u)}/{\kappa}]$.
After a straightforward algebra, we find that
\begin{equation}\label{G_solution}
\mathcal{G}(\lambda) = \frac{e^{cu_t}}{\sqrt{\cos(2\kappa u_t) +
\frac{ck_i-\tilde{\lambda}}{k_i} \frac{\sin(2\kappa u_t)}{\kappa}}} \ ,
\end{equation}
with $u_t=\frac{1}{\alpha} \ln(1+\alpha t)$.

\begin{figure}[t]
\includegraphics*[width=\columnwidth]{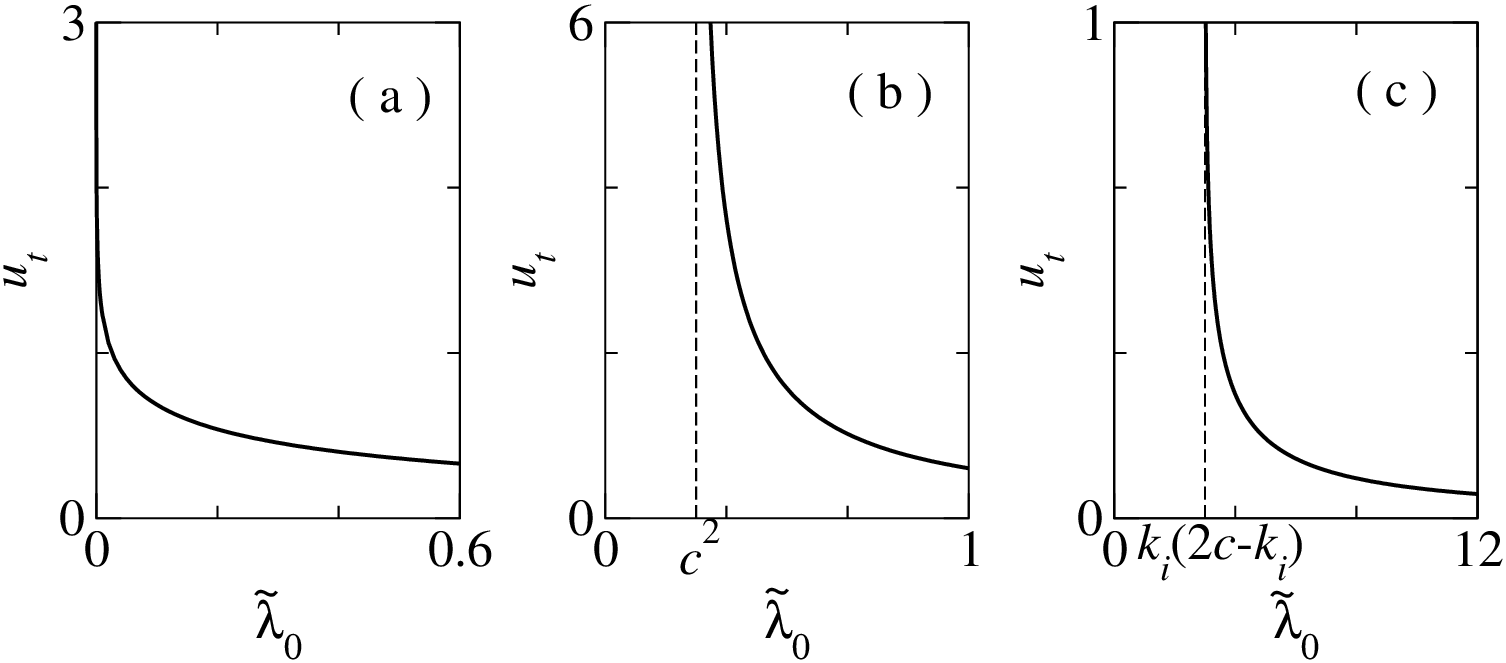}
\caption{Curves representing the relation between $u_t=\frac{1}{\alpha}\ln (1+\alpha t)$
and $\tilde\lambda_0=\alpha k_i\lambda_0 /2$
for $c<0$ in (a), $0<c<k_i$ in (b), and $k_i<c$ in (c). The
values of $(c,k_i)$ are taken to be $(-1,1)$, $(1/2,1)$, and $(2,1)$, respectively.}
\label{fig6}
\end{figure}

\begin{figure}[t]
\includegraphics*[width=\columnwidth]{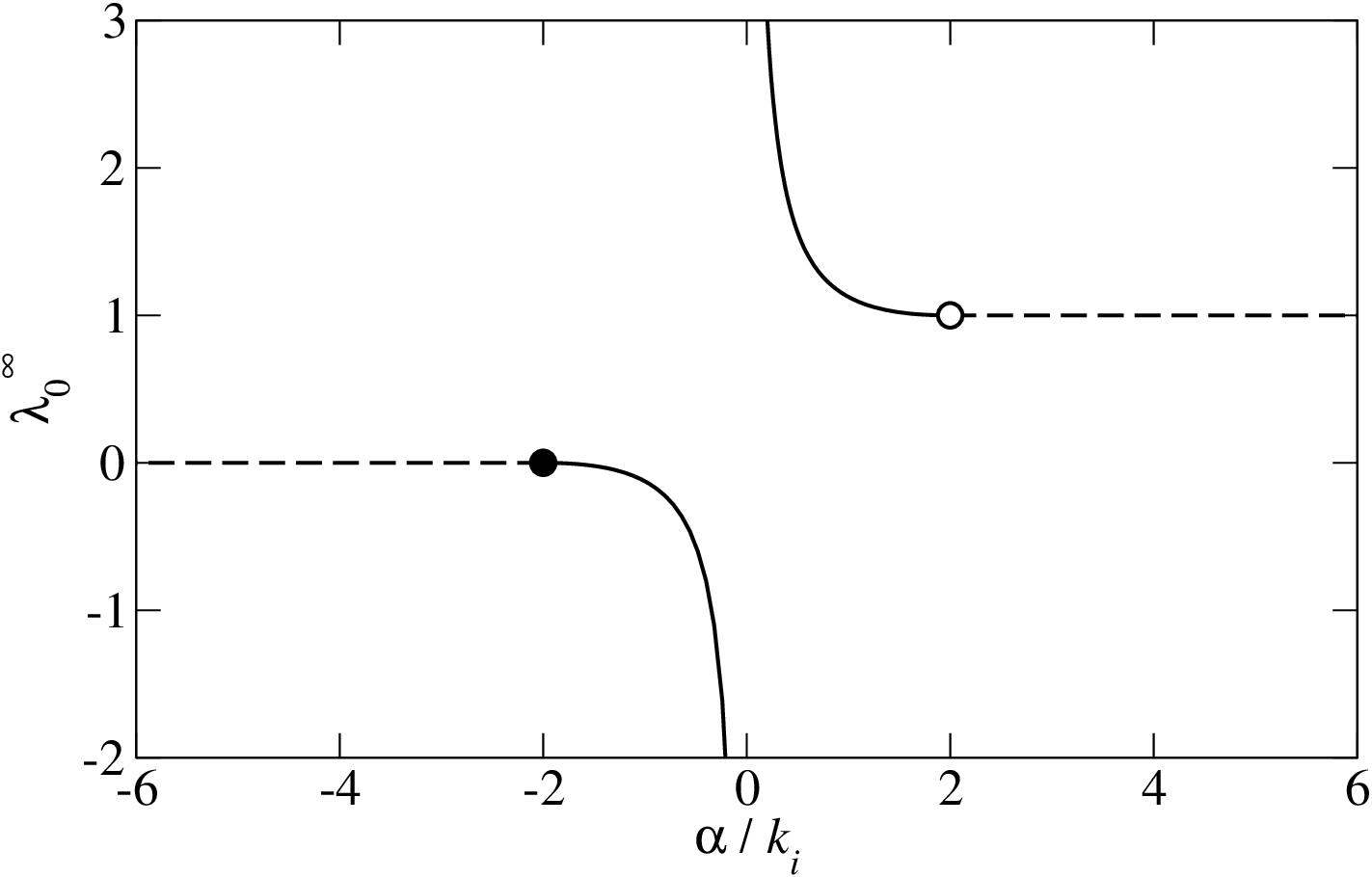}
\caption{$\lambda_0^\infty$ versus $\alpha/k_i$.
The dashed lines are a flat straight line of $\lambda_0^\infty=1$
starting from the $\alpha/k_i=2$ point (open circle), and of
$\lambda_0^\infty=0$ starting from the $\alpha/k_i=-2$ point (filled
circle).}
\label{fig7}
\end{figure}

The work PDF $P(w)$ can be obtained by the inverse Fourier transformation of
$\mathcal{G}(\lambda)$ in Eq.~(\ref{IFT}). Note that the generating function has
an inverse square-root singularity at a particular value of
$\lambda = \lambda_0(u)$ at which the denominator in Eq.~(\ref{G_solution})
vanishes.
In fact, the singularity occurs when $\tilde{A}(\tau;\lambda)=0$
or equivalently $f_\lambda(u)=0$, as seen in Eq.~(\ref{lnG_overdamped}).
The inverse square-root singularity in $\mathcal{G}(\lambda)$ at
$\lambda=\lambda_0$ implies an
exponential tail with a power-law prefactor~\cite{kwon-park-noh} in $P(w)$
\begin{equation}
P(w) \sim \frac{1}{|w|^{1/2}} e^{\lambda_0 (u) w}
\end{equation}
in the $\omega\to -\infty$ limit for $\alpha>0$ ($\lambda_0>0$) or
in the $\omega\to \infty$ limit for $\alpha<0$ ($\lambda_0<0$).

The singular point satisfies the relation
\begin{equation}\label{singular_lambda}
u_t = \frac{1}{2\sqrt{\tilde{\lambda}_0 - c^2}} \tan^{-1} \left( \frac{k_i
\sqrt{\tilde{\lambda}_0-c^2}}{\tilde{\lambda}_0 - ck_i}\right)
\end{equation}
with $\tilde{\lambda}_0 = {\alpha k_i}\lambda_0/2$. It should be
understood that $\tan^{-1}(ix) = i \tanh^{-1}(x)$ and that $0\leq \tan^{-1}{x}
< \pi$ for a real $x$.
Figure~\ref{fig6} shows the plots for the solution of
Eq.~(\ref{singular_lambda}), where the divergence of $u_t$ is observed
as $\tilde{\lambda}_0$ approaches the limiting value
from above.
Interestingly, the time dependence and the limiting value are very different
depending on whether $c<0$~($\alpha \leq -2k_i$), $0\leq c<k_i$~($-2k_i <
\alpha \leq 2 k_i$), or $k_i\geq c$~($\alpha > 2k_i$).
Especially, the limiting value
${\lambda}_0^\infty=\lim_{u\to\infty} \lambda_0$ is given by
\begin{equation}
\lambda_0^\infty = \left\{
 \begin{array}{ccc}
   0 &,& (\alpha \leq -2 k_i) \\ [2mm]
   \frac{(2k_i+\alpha)^2}{8 k_i \alpha} & , & ( -2k_i < \alpha \leq 2 k_i)
\\ [2mm]
   1 &,& (\alpha > 2 k_i)
 \end{array}\right.
 \label{lambda_0}
\end{equation}
We present the plot of the $\lambda_0^\infty$ as a function of $\alpha/k_i$ in
Fig.~\ref{fig7}.

There is an interesting symmetry of $\lambda_0^\infty(-\alpha)=
1-\lambda_0^\infty(\alpha)$.
This comes from the Crooks relation. The reverse protocol
with respect to the forward protocol $k(\tau)=k_i/(1+\alpha \tau)$
should be given as $k_r(\tau)=k(t-\tau) = k_f/(1+\alpha_r \tau)$ with
$\alpha_r=-\alpha k_f/k_i$ and $k_f = k_i/(1+\alpha t)$.
If the system starts with the EQ distribution with $k_r(0)=k_f$, all results
derived here can be also applied for the reverse process by replacing
$k_i$ by $k_f$ and $\alpha$ by $-\alpha k_f/k_i$. Then, Eq.~(\ref{lambda_0}) gives us
$\lambda_{0,r}^\infty(\alpha_r)=\lambda_0^\infty (-\alpha)$. The Crooks relation of
Eq.~(\ref{crooks-FT}) yields $\lambda_0^\infty=1-\lambda_{0,r}^\infty$ in the large $w$
limit, which leads to our symmetry of $\lambda_0^\infty(-\alpha)=1-\lambda_0^\infty(\alpha)$.

We add a few remarks on the interesting $\alpha$ dependence of $\lambda_0^\infty$:
(i)~For $\alpha \geq 2k_i$, the tail shape of $P(w)$ does not change
with $\alpha$ with fixed $\lambda_0^\infty=1$. When $\alpha$ is large enough,
the harmonic potential flattens very fast. Then, the particle dynamics
starting from the EQ distribution with $k_i$ would be rather localized
and not fully relaxed into a flattened harmonic potential.
So the fluctuation in $w$ may be
dominated by an initial transient behavior even in the long-time limit ($t\to\infty$),
independent of the detailed shape of $k(t)$. The sudden change
limit discussed in subsection~\ref{sudden} corresponds
to the $\alpha=\infty$ limit with $k_f=0$, where $\lambda_0^\infty=|k_i/(k_f-k_i)|=1$ from
Eq.~(\ref{PDF_W}) is consistent with the result for $\alpha \geq 2k_i$. Nevertheless, it is
still quite remarkable to find $\lambda_0^\infty=1$ for large but finite $\alpha$.
Similar features of initial-distribution dominance in the large deviation function
in the long-time limit have been found in various different situations~\cite{farago,visco,puglisi2,jlee}.

(ii)~For $ |\alpha| < 2k_i$, $|\lambda_\infty|$
decreases monotonically as $|\alpha|$ increases. This behavior is compatible
with the common wisdom that the fluctuation gets stronger~(longer tail in
$P(\omega)$) as the rate of the change in driving increases.


(iii) When $\alpha \leq - 2 k_i$ (or $c\leq 0$), we obtain that $\lambda_0^\infty = 0$.
This implies that $P(\omega)$ has a pure power-law tail in the positive-$w$
region in the $u\to \infty$ ($t\to 1/|\alpha|$) limit.
In this case, the driving is strong enough
to generate huge fluctuations.


Transient behavior of $\lambda_0$ is also investigated in two limits.
In the short-time limit~($u_t\to 0$),
 $P(w)$ is expected to exhibit a delta
function distribution centered at $w=0$.
This is confirmed by the solution of
Eq.~(\ref{singular_lambda}) given by $\tilde\lambda_0 \simeq k_i/(2u_t)$
or $\lambda_0 \simeq 1/(\alpha u_t)$ in the $u_t\to 0$ limit.
In the opposite limit~($u_t \to \infty$),
Eq.~(\ref{singular_lambda}) yields
\begin{equation}
\tilde\lambda_0 \simeq \left\{
 \begin{array}{ccc}
   \frac{4c^2 k_i}{k_i-2c}e^{4c u_t} &,& (c<0) \\ [2mm]
   \frac{\pi^2}{16 u_t^2} &,& (c=0) \\[2mm]
   c^2 + \frac{\pi^2}{4 u_t^2} & , & (0<c<k_i) \\ [2mm]
   c^2 + \frac{\pi^2}{16 u_t^2} & , & ( c=k_i) \\  [2mm]
   {k_i ( 2c - k_i)} + \frac{4k_i ( c-k_i)^2}{2c-k_i} e^{-4(c-k_i)u_t} &,&
                (c>k_i)
 \end{array}\right.
\end{equation}
Note that the asymptotic behavior near $\tilde\lambda=\tilde\lambda_0^+$
is very different, depending on the region. In terms of $\lambda_0$ and $t$,
it is interesting to see a nontrivial power-law relaxation for $\alpha>2k_i$ ($k_i<c$)
such that $\lambda_0\simeq 1+ z^2 (1+\alpha t)^{-z}$ with $z=1-(2k_i/\alpha)$.

The generating function also produces the cumulants of the work production by $\langle w^n\rangle_c=\left.
d^n\ln\mathcal{G}/d(-\lambda)^n\right|_{\lambda=0}$.
We focus on the mean value of the work, which is given by
\begin{equation}\label{w_mean}
\langle w\rangle =
-\frac{\alpha}{4c} \left[  k_i u_t +\frac{1}{2} \left(1-\frac{k_i}{2c}\right) ( 1 -
e^{-4c u_t} )\right]~.
\end{equation}

The quasi-static process corresponds to the limiting case where $\alpha\to 0$, $t\to\infty$
with fixed $\alpha t = (k_i-k_f)/k_f$. In this limit, we find
\begin{equation}
\langle w \rangle = -\frac{1}{2}\ln(1+\alpha t) = \frac{1}{2}\ln
\left(\frac{k_f}{k_i}\right)
\end{equation}
which agrees with Eq.~(\ref{quasi-static}).
For a sudden process, we take the opposite limit
where $\alpha\to\infty$, $t\to 0$ with the same fixed value of $\alpha t$ in the above.
Eq.~(\ref{w_mean}) approaches $\langle w\rangle = ({k_f-k_i})/{2k_i}$,
which agrees with Eq.~(\ref{sudden_cumulant}).

%

\section{discussion}\label{discussion}

The Brownian dynamics with mixture of position and momentum variables, having the even and odd parity respectively in time reversal, has not been fully scrutinized. In many literatures, the overdamped limit was taken for simplicity or as regards a light mass in experiment; otherwise the Kramers equation should be investigated, which is a nontrivial task.
Putting the position and momentum on the same footing and introducing the singular diffusion matrix as in Eq.~(\ref{singular_D}), we converted it into the usual Fokker-Planck equation,
where the strict constraint on the position
and momentum as $\delta(\dot{x}-p/m)$ is relaxed, so it is easier to handle. It is a well-known fact from the classical text books, but there are not many examples exploiting this method. In our study, we have proven this approach to be successful in finding the results analytically and numerically beyond the overdamped limit.

For the motion under the harmonic potential with a time-dependent force constant $k(t)$, we have succeeded in examining the PDF of the work production rigorously, the most important quantity in NEQ fluctuations. As a result, we have found the exponential tail with a power-law prefactor in the PDF $P(w)$ and $|\lambda_0(t)|$, the characteristic constant in the exponential tail, to decrease with time $t$ showing a fine structure of infinite but
not sharply-edged staircase. By comparing the multiple locking-unlocking  transitions (sharply-edged staircase) found in the two-dimensional linear diffusion system~\cite{noh2},
we call these rather smooth staircase as a manifestation of multiple pseudo locking-unlocking
transitions. These pseudo transitions completely go away in the overdamped limit where no rotational current exists even in the phase space.
We also consider some exactly solvable models in the overdamped limit and found an interesting power-law (not exponential) tail in $P(w)$ for the case of rather fast compression ($\alpha \le -2 k_i$) with the protocol $k(t)=k_i/(1+\alpha t)$, which implies huge NEQ fluctuations.

The potential well in optical tweezers or an optical trap experiment is controlled by an external field, so can have the shape change in time due to a time-varying external field. Therefore, our study can serve a theoretical basis for such experiments, since the potential well may be approximated to be harmonic. The perturbation theory might be exploited to investigate an anharmonic effect. Our recent study of the multi-dimensional diffusion dynamics for a linear drift force~\cite{kwon-park-noh,noh2} in the overdamped limit can also be realized in such experiments. We suggest many interesting experiments to be carried out in this direction.

\begin{acknowledgments}
We would like to thank David Thouless, Marcel den Nijs, Hong Qian, and Kyung Hyuk Kim for helpful discussions.
C. K. greatly appreciates the Condensed Matter Group at University of Washington (UW) for the support during his sabbatical year at UW. This work was supported by Mid-career Researcher Program through NRF grant (No.~2010-0026627) funded by the MEST.
\end{acknowledgments}


\begin{thebibliography}{99}
\bibitem{exp1} The stationary distribution for general muti-dimensional linear diffusion systems
with positive definite $\mathsf{D}$ ($\epsilon>0$) can be found in~\cite{kwon-park-noh,gardiner,risken}.
\bibitem{exp2} A similar property was found also in two-dimensional linear diffusion systems in the
overdamped limit~\cite{kwon-park-noh}.
\end{thebibliography}
\end{document}